\documentclass[a4paper,pra,superscriptaddress,11pt,showkeys]{revtex4}
\usepackage[a4paper,centering,hmargin=1.75cm,vmargin={2cm,3.6cm}]{geometry}
\usepackage{graphicx}

\usepackage{amsmath}
\usepackage{amsfonts}
\usepackage{amssymb}

\begin{document}

\title{Silicon nitride on-chip C-band spontaneous emission generation based on~lanthanide doped microparticles}

\author{Dmitry~V.~Obydennov}
\email{obydennovdv@my.msu.ru}
\affiliation{Institute of Nanotechnology of Microelectronics of the Russian Academy of Sciences, Moscow 119334, Russia}

\author{Ilya~M.~Asharchuk}
\affiliation{Institute of Nanotechnology of Microelectronics of the Russian Academy of Sciences, Moscow 119334, Russia}

\author{Alexander~M.~Mumlyakov}
\affiliation{Institute of Nanotechnology of Microelectronics of the Russian Academy of Sciences, Moscow 119334, Russia}

\author{Maxim~V.~Shibalov}
\affiliation{Institute of Nanotechnology of Microelectronics of the Russian Academy of Sciences, Moscow 119334, Russia}

\author{Nikolay~A.~Vovk}
\affiliation{Institute of Nanotechnology of Microelectronics of the Russian Academy of Sciences, Moscow 119334, Russia}

\author{Ivan~A.~Filippov}
\affiliation{Institute of Nanotechnology of Microelectronics of the Russian Academy of Sciences, Moscow 119334, Russia}

\author{Lidiya~S.~Volkova}
\affiliation{Institute of Nanotechnology of Microelectronics of the Russian Academy of Sciences, Moscow 119334, Russia}

\author{Michael A. Tarkhov}
\affiliation{Institute of Nanotechnology of Microelectronics of the Russian Academy of Sciences, Moscow 119334, Russia}

\begin{abstract}
    The integration of active light-emitting elements into planar photonic circuits on a silicon nitride platform remains challenging due to material incompatibilities and high-temperature processing. Proposed hybrid method embeds monodisperse luminescent particles into lithographically defined wells above a 200 nm-thick silicon nitride taper coupler. A fabrication process involving wells etching, particle deposition, and planarization enables precise integration while maintaining waveguide integrity. When pumped at 950 nm with a diode laser, the device emits broadband radiation in the 1500–1600 nm range, covering the optical telecommunication C-band. Numerical simulations yield an average coupling efficiency of 0.25\% into the fundamental waveguide mode, suggesting significant potential for further device optimization. The approach provides a scalable route for integrating broadband telecommunications emitters on a silicon nitride platform.
\end{abstract}


\maketitle

\section{Introduction}
Nowadays, research and development in the field of active and passive elements for planar photonics remain highly relevant~\cite{xiang_silicon_2022,tian_programmable_2023}, driven by significant technological breakthroughs in the fabrication of high-quality materials. One of the promising platforms for planar photonics is silicon nitride: this material, paired with silicon dioxide, enables low intrinsic losses~\cite{gardes_review_2022,guo_characterization_2004, tien_ultra-low_2010} compared to other widely used platforms such as InP~\cite{augustin_inp-based_2018,smit_past_2019,van_der_tol_inp-based_2010}, GaAs~\cite{koester_gaas-based_2024}, and SOI~\cite{yan_monolithic_2021}. The fabrication technology for silicon nitride-based planar photonic elements is well-established and has been implemented in a variety of devices~\cite{romero-garcia_silicon_2013,sacher_multilayer_2015, xu_frequencydependent_2024, buzaverov_silicon_2024, belogolovskii_large_2025, psiquantum_team_manufacturable_2025}. However, a critical technological challenge persists in integrating active elements such as photodiodes, amplifiers, and photodetectors into silicon nitride-based planar photonic chips. This challenge arises primarily from the incompatibility of silicon nitride with active elements operating in the near-infrared telecommunications band (C-band) and from the high-temperature annealing processes ($\sim1300^\circ$C) required during waveguide fabrication. Current solutions for integrating active elements into silicon nitride planar optical structures typically rely on hybrid assembly methods~\cite{zhu_loss_2018}.

In this work, we propose a technology for creating a C-band spontaneous emission generator integrated into a planar waveguide, operating on the principle of light down-conversion. The active material consists of luminescent particles based on NaYF$_4$ doped with erbium and ytterbium. Recently, such particles have found broad applications in bioimaging, biodiagnostics~\cite{oleksa_polynn-dimethylacrylamide-coated_2021,cui_synthesis_2016, chen_controllable_2011, reineck_nearinfrared_2017}, chemical sensing~\cite{abbasi-moayed_application_2020}, and in creating new gain media in planar photonic amplifiers~\cite{zhang_high-gain_2016, zhou_optical_2021, yang_great_2020} and lasers~\cite{sun_ultralarge_2022, moon_continuous-wave_2021}. These particles exhibit a broad emission spectrum: upon absorbing light in the 900–1000 nm range~\cite{wang_solvothermal_2010, gunaseelan_reverse_2020}, they emit in the visible and near-infrared (C-band, 1530–1565 nm) regions, attributed to the rich energy structure of erbium ions~\cite{nadort_lanthanide_2016,zhai_enhancement_2013}. While there is an extensive class of infrared luminescent materials~\cite{zhang_ultra-broadband_2023,chen_broadening_2023,gong_novel_2025,gao_research_2023,rajendran_invited_2019}, their integration using microelectronic standards presents significant challenges. The NaYF$_4$:Yb,Er particles employed here are usually monodisperse in size, which is crucial for precise integration. Recent work by Asharchuk~\textit{et al.}~\cite{asharchuk_planar_2024} demonstrated the feasibility of large-scale ``printing'' of these particles on substrates up to 100 mm in diameter.

This approach enables the precise placement of luminescent microparticles at designated wells above nanophotonic structures fabricated in silicon nitride. To date, few attempts have been made to combine silicon nitride platforms with such particles due to the technological complexity of integration. In Ref.~\cite{liu_gain_2022}, a silicon nitride slot waveguide filled with PMMA containing fluoride particles was proposed to enhance active medium and waveguide coupling. Here, we present a design based on a circular-sector-shaped taper with luminescent particles deposited above. We have fabricated this device and demonstrated its functionality by coupling of particle emission into the waveguide.

\section{Results and discussion}

\subsection{Luminescent properties of particles}

\begin{figure}[htb!]
\centering
\includegraphics[]{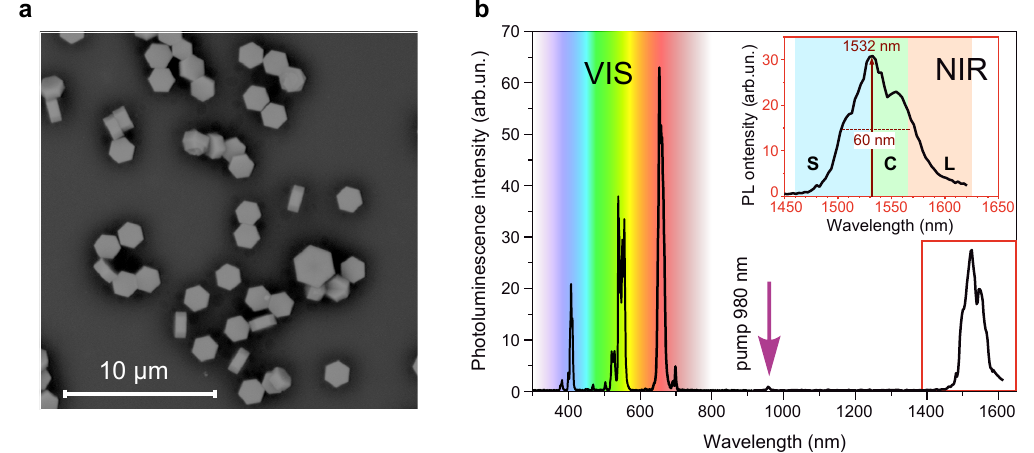}
\caption{Luminescence of the individual particles. (a) Scanning electron micrograph of particles deposited on a silicon substrate for luminescence studies. (b) Measured photoluminescence spectrum of the luminescent particles. The inset shows the near-infrared spectral region, highlighting the emission peak and linewidth. Optical telecommunication bands are denoted by the letters 'S', 'C', and 'L', and highlighted by blue, green, and pink color boxes correspondingly.}
\label{spectrum}
\end{figure}

We started with examination of the luminescent characteristics of the synthesized monodisperse \\NaYF$_4$:Yb$^{3+}$, Er$^{3+}$ microparticles fabricated using the methodology described in   \textit{Asharchuk~et al.}~\cite{asharchuk_planar_2024}. A scanning electron micrograph (SEM) of the particles deposited on a silicon substrate is shown in Figure~\ref{spectrum}a. Luminescence spectroscopy was performed on individual particles;
Figure~\ref{spectrum}b displays the luminescence spectrum of the NaYF$_4$:Er,Yb, particles under 975 nm laser excitation. One can note from the plot that these particles exhibit intense emission bands in both the visible and infrared regions; in the present work, we focus on their near-infrared emission. The peak emission intensity in the infrared regime occurs at 1532 nm, with a full width at half maximum (FWHM) of 60 nm, as illustrated in the inset of Figure~\ref{spectrum}b.

\subsection{Chip fabrication}
\begin{figure}[htb!]
    \centering
    \includegraphics[]{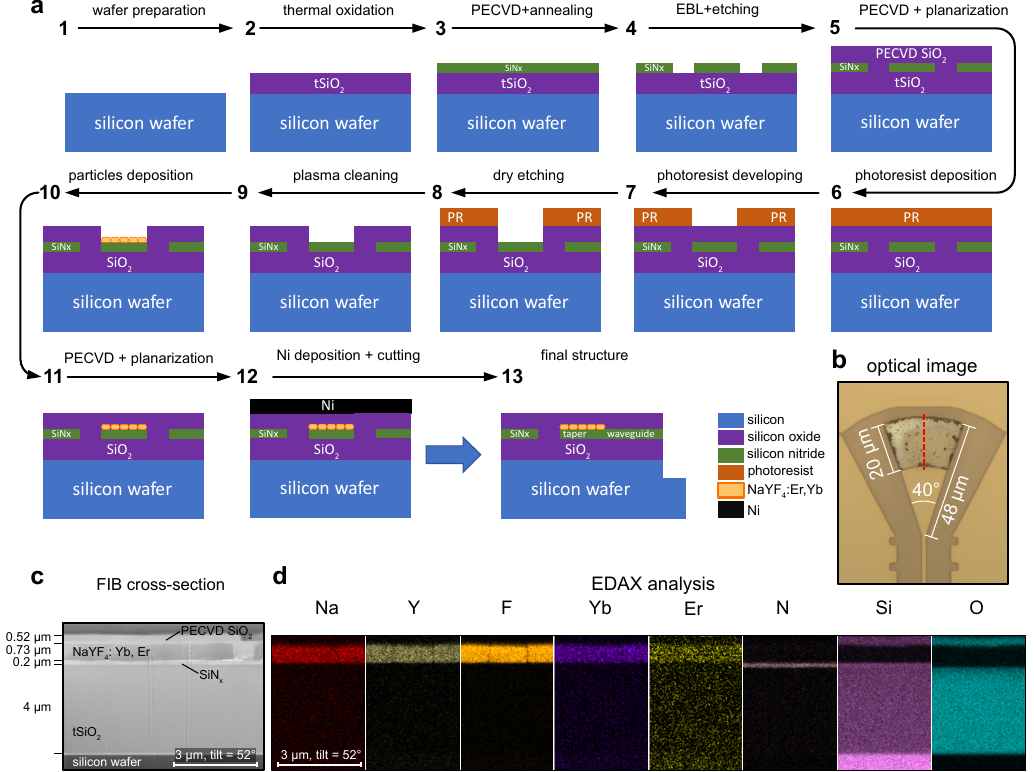}
    \caption{Fabrication of the chip with luminescent particles. a) Fabrication workflow. 1. Preparation of the silicon wafer for thermal oxidation; 2. wafer thermal oxidation in water vapor; 3. silicon nitride deposition and annealing; 4. EBL patterning and dry etching of the waveguide and coupler; 5. PECVD deposition and planarization of an upper SiO$_2$ layer; 6. photoresist deposition and particles well region formation; 7. photeresist development; 8. SiO$_2$ dry etching; 9. plasma removing of the photoresist; 10. filling the well region with particles according to~\cite{asharchuk_planar_2024}; 11.PECVD and planarization of an upper SiO$_2$ layer; 12. nickel mask deposition for deep etching of the multilayer structure to form the chip facet; 13. schematic cross‐section of the chip, showing its internal structure. b) Optical micrograph of the resulting planar taper structure with deposited microparticles. The red dashed line indicates the path of the focused ion beam cutting. c) SEM cross‐section of the structure with layer indications. d) Energy‐dispersive X‐ray spectroscopy (EDAX) maps corresponding to the elemental peaks.}
    \label{chip}
\end{figure}

The photon chip designed to couple the luminescence emission from microparticles with the waveguide composed of a taper of a circular sector shape with a radius of 48 $\mu$m and an opening angle of 40$^\circ$, which transitioned into  1.4 $\mu$m-width waveguide. Luminescent particles NaYF$_4$:Yb$^{3+}$,Er$^{3+}$ were deposited on the rear area of the taper, as it can be seen in Figure~\ref{chip}b. The sample was fabricated on a silicon wafer.  The fabrication process, similar to that described in Mumlyakov et al. \cite{mumlyakov_void-free_2024}, included thermal oxidation of silicon to form a silicon dioxide layer, deposition of a silicon nitride layer, patterning of the waveguide and taper from the silicon nitride layer, passivation with silicon dioxide, and incorporation of particles atop the lithographic pattern. The lateral dimensions of most particles lie in range of 1.5–2 $\mu$m, with heights of 0.7–0.9 $\mu$m \cite{asharchuk_planar_2024}; therefore, the particle well area depth was set to 1 $\mu$m. Within the well region, particles were distributed upon deposition according to the hexagonal close‐packing principle. To increase output emission collection efficiency, a 500 $\mu$m‐long taper narrowing to 0.5 $\mu$m at the chip facet was formed at the end of the waveguide. Figure~\ref{chip}a presents the sequence of 13 primary technological steps, demonstrating the process of fabricating planar waveguides and couplers with integrated microparticles. The manufacturing steps were performed in the following order: preparation of the silicon wafer for thermal oxidation; thermal oxidation of the wafer in water vapor at 950$^\circ$C, achieving a 4 $\mu$m thick silicon dioxide layer; deposition of a 200 nm silicon nitride film by plasma‐enhanced chemical vapor deposition (PECVD) followed by annealing at 1150$^\circ$C for 8 hours; patterning of the waveguide and coupler using electron‐beam lithography (EBL) followed by reactive ion etching in a C$_4$F$_8$/Ar gas mixture; deposition of an upper silicon dioxide buffer layer by PECVD followed by planarization; application of photoresist and foration of the well region for NaYF$_4$:Er,Yb microparticles; development of the photoresist; dry etching of silicon dioxide to a depth of 1 $\mu$m to form the well region; plasma removing of the photoresist; filling of the well region with particles according to the method described in \cite{asharchuk_planar_2024}; deposition of silicon dioxide by PECVD for particle passivation followed by planarization; deposition of a nickel hardmask for deep etching of the multilayer structure to form the chip facet. After facet formation, the nickel layer was removed.

After fabricating the chip and performing all optical measurements, described below, we examined the internal structure of the sample by analyzing a cross‐section created via focused ion beam milling, followed by investigation of the coupler region using scanning electron microscopy and energy‐dispersive X‐ray spectroscopy (EDAX). The results are presented in Figure~\ref{chip}c,d. As seen in the figure, the elemental composition of the cross‐section layers matches the design: a 4 $\mu$m silicon dioxide layer, a 200 nm silicon nitride layer, a 730 nm NaYF$_4$:Yb,Er particle layer, and a 0.52 $\mu$m silicon dioxide layer.

\subsection{Optical measurements and calculations}
\begin{figure}[htb!]
     \centering
     \includegraphics[]{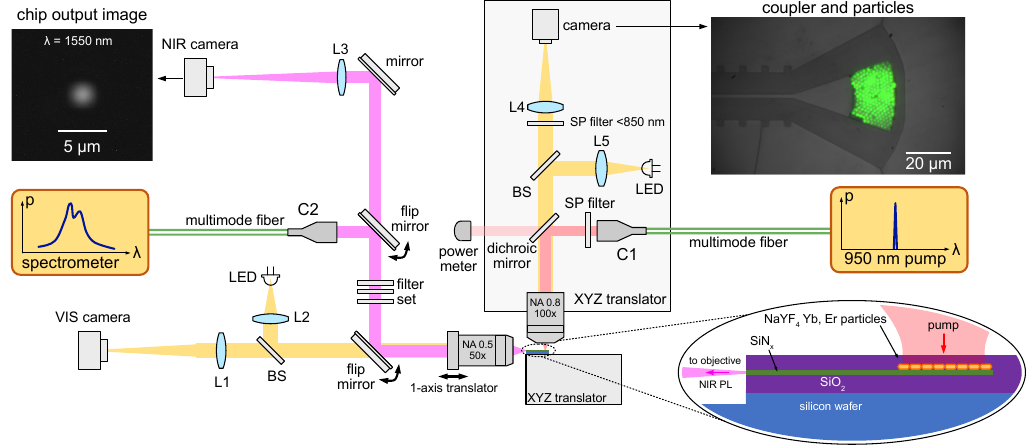}
     \caption{Experimental setup. L1–L5: lenses; C1, C2: fiber collimators; LED --- illumination; BS ---beam splitter; SP filter --- short-pass filter. Insets: the optical micrograph of the taper with visible upconversion emission taken through the pump arm, the NIR image of the emission spot at the chip output, and the illustration of the pump and collection geometry.}
     \label{setup}
\end{figure}

After fabrication of the photonic chip, we measured the power of the luminescent emission from the particles that coupled into the waveguide after propagating through the coupler. The measurement setup scheme is shown in Figure~\ref{setup}. A fiber‐coupled semiconductor laser diode emitting at a wavelength of 950 nm was employed for particle excitation. The excitation system comprised a fiber collimator C1, an optical filter to suppress the diode’s satellite lines, a dichroic mirror, and a focusing objective with numerical aperture NA = 0.8. The excitation spot on the sample was 30 µm in diameter, fully covering all particles deposited on the taper. Emission coupled into the waveguide was collected at the opposite chip facet using an objective with NA = 0.5. After that, the collected emission passed through a filter set to remove residual pump light and was directed either to an infrared camera (SWIR1300KMA, ToupTek, China) or into fiber collimator C2, which was connected to a spectrometer (Optosky, China).

\begin{figure}[htb!]
     \centering
     \includegraphics[]{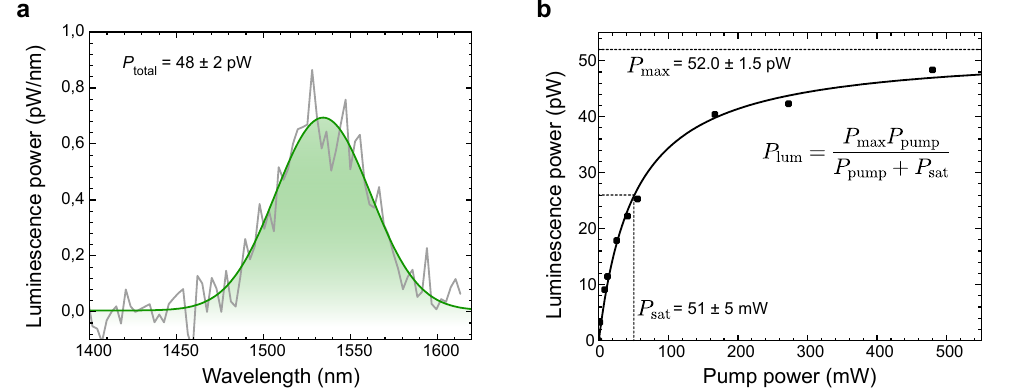}
     \caption{Emission coupling into the waveguide. a) Spectral power of luminescent emission at the chip output. Gray curve: experimental data; green bell: Gaussian fit. $P_{\mathrm{total}}$ denotes the integrated power over the specified range. b) Integrated luminescence power at the chip output as a function of excitation power. Markers: experimental data; solid curve: saturation‐law fit. $P_{\max}$ and $P_{\mathrm{sat}}$ denote maximum luminescence power and saturation power, respectively.}
     \label{lum}
\end{figure}

We recorded emission power spectra transmitted through the waveguide and collection optics at various pump powers. The results are presented in Figure~\ref{lum}. As seen in Figure~\ref{lum}a, the peak spectral power reaches 0.8 pW/nm at approximately 1530 nm, the spectral full width at half maximum is  $\sim 60$ nm, and the total integrated power is $48 \pm 2$ pW according to the results of peak approximation. The dependence of luminescence power on pump power, shown in Figure~\ref{lum}b, exhibits the characteristic saturation behavior of down‐conversion luminescence. The data were fitted to the saturation model $P_{\mathrm{lum}}(P_{\mathrm{pump}}) = P_{\max}\,P_\mathrm{pump}/\left( P_\mathrm{pump} + P_\mathrm{sat}\right),$ where $P_{\mathrm{lum}}$ is the luminescence power, $P_{\mathrm{pump}}$ the pump power, $P_{\max}$ the maximal luminescence power, and $P_{\mathrm{sat}}$ the saturation pump power. We obtained $P_{\max} = 52.0 \pm 1.5$ pW and $P_{\mathrm{sat}} = 51 \pm 5$ mW, corresponding to a pump intensity of $7.2 \times 10^{3}$ W/cm$^{2}$.

\begin{figure}[htb!]
     \centering
     \includegraphics[]{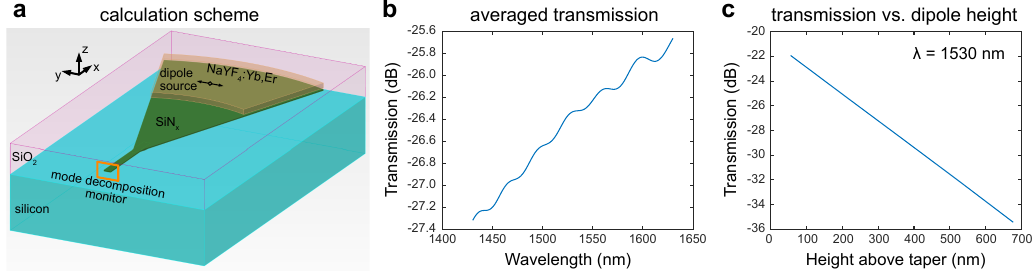}
     \caption{Simulation of particle emission coupling into the waveguide mode. a) Simulation schematic. b) Transmission from a dipole source into the TE mode, averaged over the luminescent particle positions shown in (a). c) Averaged transmission as a function of dipole source height.}
     \label{calc}
\end{figure}

To evaluate the coupling efficiency of the device, we performed numerical simulations of emission coupling into the waveguide mode using the finite‐difference time‐domain (FDTD) method. The luminescent particles were modeled as incoherent dipole sources. Given the excitation geometry, we assume the dipole moments lie predominantly in the prism base plane of the particles \cite{green_nanoplasmonic_2017}. Thus, the main contribution arises from dipoles oriented in the sample plane, perpendicular to the waveguide, coupling primarily into the TE mode. In our simulations, only this dipole polarization and the TE mode were considered. The simulated structure geometry matched that of the fabricated chip; the simulation schematic is shown in Figure~\ref{calc}a. The dipole source was scanned across the particle coordinates, and transmission into fundamental TE mode was collected. The resulting transmission of dipole emission into the fundamental TE waveguide mode, averaged over all source positions, is plotted in Figure~\ref{calc}b. The mean transmission is approximately $-26$ dB (0.25\%), with the highest contributions from sources nearest to the taper. To illustrate this, we averaged transmission over $x,y$ coordinates to obtain its dependence on dipole height. The result, shown in Figure~\ref{calc}c, reveals an exponential decay of coupling efficiency with increasing distance between the dipole and the structure surface. This observation highlights the general challenge of efficient emission coupling into a waveguide: significant spatial overlap between the particle emission field and the guided mode is required.

To further enhance device performance, several optimization strategies can be considered. Tailoring the taper geometry, for instance, by adjusting the opening angle or introducing a multi-stage taper, could improve spatial mode matching and increase coupling efficiency. Improving the passivation technology could decrease the scattering losses, induced by surface roughness. Engineering of the particle distribution, including controlled variation of particle density and orientation, can optimize local dipole alignment relative to the waveguide modes. Finally, integrating a resonant cavity or photonic crystal structure in the taper region may provide Purcell-enhanced emission and directional coupling.

\section*{Conclusion}
In this work, we demonstrate the experimental realization of a planar integrated spontaneous emission generator in the C‑band based on monodisperse microparticles NaYF$_4$:Yb,Er embedded in a silicon nitride photonic structure. The developed particle deposition and passivation technology enables reliable fixation of the particles over the coupler, ensuring their uniform distribution and optical coupling to the fundamental TE–mode. Experimentally, we show that pumping at 950~nm with a semiconductor laser successfully couples intense emission into the waveguide, with a peak at 1530~nm, a spectral width of 60~nm (FWHM), and an integrated power of $48\pm2$~pW at the chip output. Numerical FDTD simulations demonstrate that the coupling efficiency of upconversion dipole emitters into the TE–mode is on the order of 0.25\%. Further optimization of the structural parameters has the potential to yield more efficient spontaneous emission generators. The proposed approach opens prospects for creating active elements on the silicon nitride platform without complex hybrid assemblies: integrated microparticles can serve as broadband emission sources in the telecommunications range, amplify signals in planar amplifiers, and act as the gain medium for microlasers.

\subsection*{Acknowledgements}
The study has been carried out with the support of project No. 125020501540-9 of the Ministry of Education and Science of the Russian Federation. Fabrication and technology characterization were carried out at large scale facility complex for heterogeneous integration technologies and silicon + carbon nanotechnologies.
\subsection*{Conflict of Interests}
The authors declare no conflict of interest.

\bibliographystyle{ieeetr}
\bibliography{main}
\end{document}